\documentclass[journal=jcim,manuscript=article]{achemso}

\usepackage[version=3]{mhchem} 
\usepackage{hyperref}
\usepackage{xcolor}
\usepackage{subcaption}

\author{Ivan Žugec}
\email{zugec.ivan@gmail.com}
\affiliation[CFM/CSIC-UPV/EHU]{Centro de Física de Materiales CFM/MPC, CSIC-UPV/EHU, Paseo Manuel de Lardizabal 5, 20018, Donostia-San Sebastián, Spain}
\alsoaffiliation[UPV/EHU]{Departamento de Polímetros y Materiales Avanzados: Física, Química y Tecnología, Facultad de Química (UPV/EHU), Apartado 1072, Donostia-San Sebastián, 20080, Spain}

\author{Tin Hadži Veljković}
\affiliation[UvA]{UvA-Bosch Delta Lab, University of Amsterdam, Amsterdam Science Park 904, Amsterdam, 1098 XH, Netherlands}

\author{Maite Alducin}
\email{maite.alducin@ehu.eus}
\affiliation[CFM/CSIC-UPV/EHU]{Centro de Física de Materiales CFM/MPC, CSIC-UPV/EHU, Paseo Manuel de Lardizabal 5, 20018, Donostia-San Sebastián, Spain}
\alsoaffiliation[DIPC]{Donostia International Physics Center, Paseo Manuel de Lardizabal 4, Donostia-San Sebastián, 20018, Spain}

\author{J. Iñaki Juaristi}
\email{josebainaki.juaristi@ehu.eus}
\affiliation[UPV/EHU]{Departamento de Polímetros y Materiales Avanzados: Física, Química y Tecnología, Facultad de Química (UPV/EHU), Apartado 1072, Donostia-San Sebastián, 20080, Spain}
\alsoaffiliation[CFM/CSIC-UPV/EHU]{Centro de Física de Materiales CFM/MPC, CSIC-UPV/EHU, Paseo Manuel de Lardizabal 5, 20018, Donostia-San Sebastián, Spain}
\alsoaffiliation[DIPC]{Donostia International Physics Center, Paseo Manuel de Lardizabal 4, Donostia-San Sebastián, 20018, Spain}

\title[An \textsf{achemso} demo]
  {Dynamic Training Enhances Machine Learning Potentials for Long-Lasting Molecular Dynamics}


\begin{document}


\begin{abstract}
  Molecular Dynamics (MD) simulations are vital for exploring complex systems in computational physics and chemistry. While machine learning methods dramatically reduce computational costs relative to ab initio methods, their accuracy in long-lasting simulations remains limited. Here we propose dynamic training (DT), a method designed to enhance accuracy of a model over extended MD simulations. Applying DT to an equivariant graph neural network (EGNN) on the challenging system of a hydrogen molecule interacting with a palladium cluster anchored to a graphene vacancy demonstrates a superior prediction accuracy compared to conventional approaches. Crucially, the DT architecture-independent design ensures its applicability across diverse machine learning potentials, making it a practical tool for advancing MD simulations.
\end{abstract}

\section{Introduction} \label{sec1}
Molecular dynamics (MD) simulations have proven to be a powerful and versatile tool, providing valuable insights into the mechanisms behind complex phenomena \cite{rahman1964correlations,karplus2002molecular,schlick2010molecular}.
Moreover, MD simulations also hold significant promise in optimizing and accelerating discovery of novel materials \cite{pilania2022recent}.  
However, with current MD approaches, one often needs to choose between accuracy and simulation speed. Ab initio methods such as density functional theory (DFT) provide highly accurate predictions but are computationally expensive and scale poorly with system size. In contrast, classical force fields offer near-linear scaling with system size, but often lack accuracy and the transferability required for application to diverse systems. In recent years machine learning potentials (MLPs) have emerged as a powerful alternative to the aforementioned methods as they offer the possibility of achieving accuracy comparable to that of ab initio methods, while maintaining near-linear scaling with system size due to their predominantly local nature. MLPs come in various forms, including kernel methods \cite{rupp2012fast,chmiela2017machine,dral2019mlatom}, permutationally invariant polynomials \cite{qu2018permutationally, houston2023pespip}, and neural networks
\cite{behler2007generalized, gilmer2017neural, smith2017ani, schutt2018schnet, zhang2021, zhang2022, gasteiger2021gemnet, batzner20223, batatia2022mace}. Among these, neural network potentials (NNPs) have emerged as a particularly promising avenue for creation of accurate and efficient multidimensional potential energy surfaces (PES)  \cite{shakouri2017, binjiang2020,  Zeng2023, deng2023chgnet, musaelian2023learning, zugec2024understanding,  Muzas2024, vzugec2024global, bochkarev2024graph, Omranpour2025}. However, the application of neural network potentials is not without challenges. The requirement for extensive training data can be a substantial barrier, particularly for systems for which high-quality reference calculations are computationally expensive. Furthermore, the standard training paradigm, which focuses on minimizing the global error in single-step predictions, may not adequately capture local intricacies present in the PES. This disparity becomes apparent in applications such as MD simulations, where the cumulative effect of errors and  
exposure to varying temperatures can drive the system into regions where the potential is less accurately learned, frequently causing instabilities in the simulated dynamics \cite{fu2022forces}. 

In this work, we propose a dynamic training (DT) approach for enhancing the training of NNPs from ab-initio molecular dynamics (AIMD) simulations. We apply this strategy to an equivariant graph neural network (EGNN), resulting in what we term DT-EGNN. In contrast to conventional approaches that process data points in isolation, our method explicitly accounts for the sequential nature of MD simulations by including integration of equations of motion into the training process of a neural network. Thus, it enables direct comparison between predicted simulations and AIMD reference data,
enhancing the ability of a model to capture the temporal evolution of the system. To exemplify this statement, a comprehensive and challenging dataset of AIMD simulations describing the dynamics of $\text{H}_2$ molecules interacting with $\text{Pd}_6$ clusters anchored in graphene vacancies (H$_2$/Pd$_6$@G$_{\text{vac}}$) \cite{alducin2019dynamics} is used and we demonstrate that DT-EGNN achieves higher accuracy compared to conventional training methods while indicating promising data efficiency. 

\section{Results}\label{sec2}
\subsection{Method Description}
\begin{figure} 
    \centering
    \includegraphics[width=0.75\textwidth]{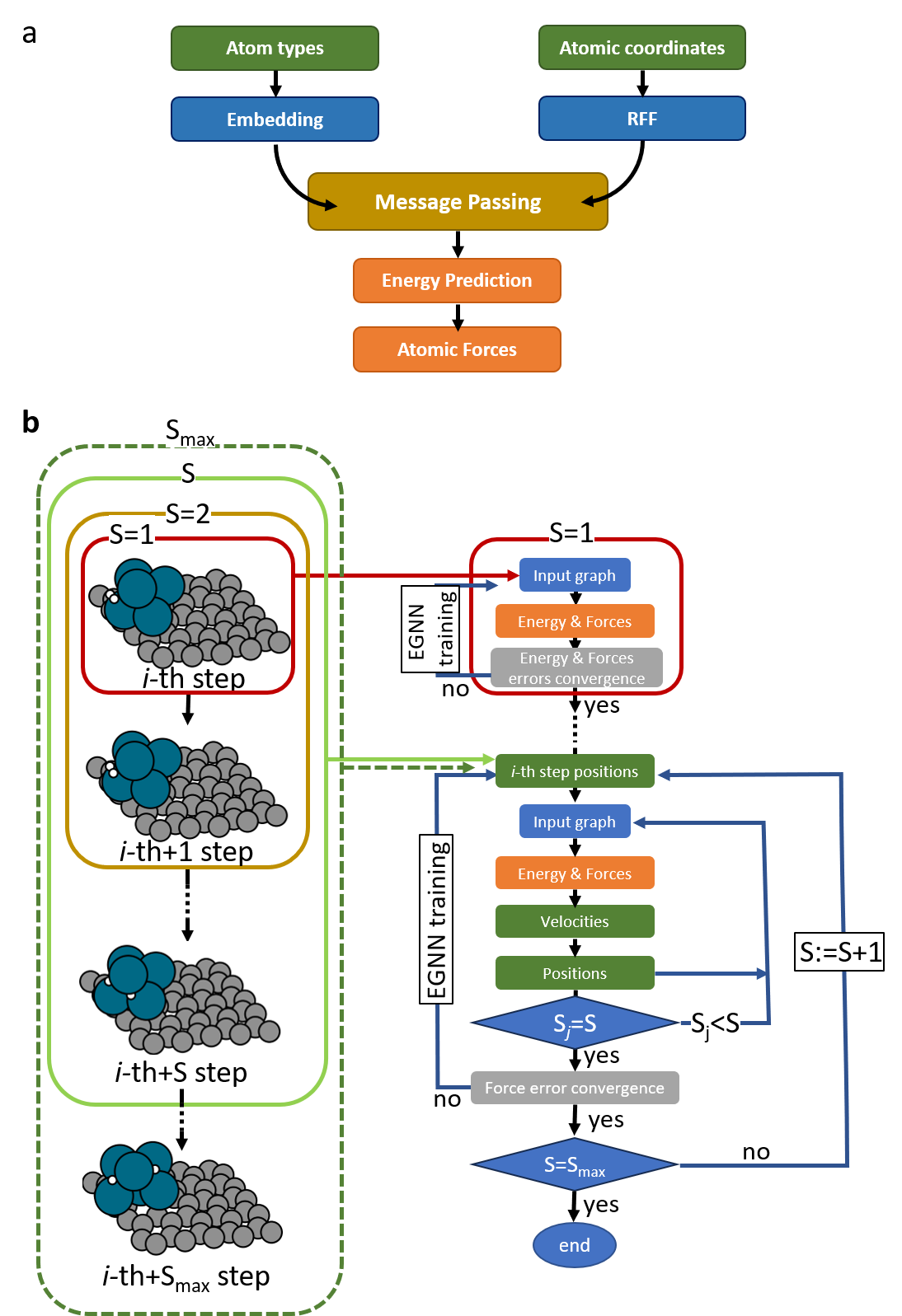}
    \caption{\textbf{DT-EGNN method. a} Schematic representation of the EGNN architecture used in this work. Atom types undergo processing through the embedding layer while atomic coordinates are mapped to random Fourier features (RFF). These inputs are then updated via message passing layers, leading to energy prediction. Finally, atomic forces are computed as the negative gradient of the energy. \textbf{b} Schematic representation of the DT method. It starts by training a model on all the initial structures present in the training points ($S=1$). Once the convergence criteria is met, the dynamics information is progressively expanded by increasing the subsequence length $S$ by one. In general, training for a given $S$ starts by predicting the atomic forces for the initial structures that, together with positions and velocities, determine the next-step atomic coordinates. These coordinates are mapped to a new input graph, enabling prediction of atomic forces and corresponding velocities. The loop continues until the desired subsequence length is reached.}
    \label{fig:flowcharts}
\end{figure}

Development of machine learning models in computational chemistry is often hindered by the scarcity and high computational cost of accurate data. One way to speed up the process of creating a dataset is to use AIMD. Many widely-used datasets, including MD17 \cite{chmiela2017machine}, Open Catalyst Project \cite{chanussot2021open}, and ANI-1x \cite{smith2020ani} are either partially or fully generated from AIMD. The common way of utilizing these datasets is to randomly select atomic structures for training, validation, and testing. While this approach simplifies data handling, it discards valuable temporal information present in the simulations. Given that many NNP applications revolve around performing MD simulations, it is reasonable to expect that incorporating temporal structure within the training process would enhance the NNP quality. Therefore, in this work, we propose treating each data point as a subsequence of an AIMD simulation rather than as an isolated atomic configuration. This in turn enables us to incorporate molecular dynamics directly into the training process of a NNP.  In order to implement this approach, we have chosen an EGNN  as our NNP architecture (see Figure~\ref{fig:flowcharts}\textbf{a}). Detailed information about the EGNN architecture employed in this work is provided in the Methods section.

Starting with the data preprocessing step we extract the information about the unit cell, types of atoms, and their positions for each atomic structure in the dataset. With these quantities we can form a graph for each atomic structure. Node features of a graph are represented by a one-hot vector representing the atom element, whereas the edge features of the graph are represented by the distances between atoms. Which nodes are connected, i.e., the so-called node neighborhood, is determined by the radius graph method. Global graph features we are aiming to learn are represented by the DFT computed energy and atomic forces for a given atomic structure. However, barring practical considerations such as memory consumption, there are no limitations on how much information we can record into the data structure. It is therefore at this point that we leverage the sequential nature of the AIMD data. Each data point, be it in the training or validation set, has information on not only the DFT calculated energy and forces for the given atomic structure, but also the atomic forces for the ensuing $S_{\text{max}} - 1$ atomic structures that follow it in the corresponding AIMD simulation. Here, $S_{\text{max}}$ is the predefined upper limit for the subsequence length and can, in general, be different for training and validation points. If the given atomic structure happens to be at the $i$-th timestep of an AIMD simulation, we take atomic force information of the next $[i + 1, \dots, i + S_{\text{max}} - 1]$ atomic configurations calculated in AIMD. Finally, together with the integration timestep, we also store the atomic positions and velocities of the $i$-th structure because they provide initial conditions for the dynamics of the subsequence that takes place in the training process.
This approach naturally extends to the points in the validation set as they also become subsequences of AIMD simulations. While forward passes require storing gradients for neural network optimization, validation passes do not have this memory constraint, enabling us to use subsequences that are an order of magnitude longer than those in the training set.

In order to perform stable dynamics within the forward pass, we must ensure that the predictions on initial atomic structures are as accurate as possible. To achieve this, the training process starts with the standard practice of minimizing prediction errors of energies and forces on single atomic structures, which correspond to the initial atomic structure of each training point.  It is useful to frame them as subsequences of length one ($S=1$). Once the convergence criteria is met, instead of terminating the training process, we continue with the training by incrementing the subsequence length by one. Consequently, this makes the training iteration to consist of more parts as summarized in Figure~\ref{fig:flowcharts}\textbf{b}.
Similarly to $S=1$, it starts with the prediction of the energy and forces for the initial atomic structures. These forces, in conjunction with velocities and positions, which were stored in the data preprocessing part, are utilized to derive the next-step atomic coordinates. These coordinates can now be used to build a graph in a similar way as described before. Upon generation of a new graph, the model predicts new atomic forces, which are then combined with the forces from the previous step to update the velocities using the velocity Verlet algorithm. This forms a loop that occurs once if the subsequence length is two, and more generally $S-1$ times if the length of the subsequence is $S$. Once the model yields the predictions of energies and atomic forces for the whole subsequence, they are compared to those obtained in the corresponding AIMD calculations that were recorded during data preprocessing step.  The loss function, as well as other important details regarding the training process are described in the Methods section.

Notice that the error between the model prediction and the corresponding DFT calculation of any given structure within the subsequence will depend on all the model predictions that came before it. This inevitably comes from the dynamic nature of the training process. Furthermore, all predictions are connected in a computational graph through which gradients can flow. This statement is not trivial as we show shortly, but it means that the network will be penalized for predictions that lead to high errors as simulation progresses. Another way to think of it is that subsequences act as a type of regularization that pushes the network weights to local minima more suitable for the task of performing long-lasting dynamic simulations.

Here, a remark regarding the calculation of the atomic neighborhoods in DT is in order. For subsequence lengths greater than one, each update of the atomic positions during the dynamical training can modify the underlying atomic neighborhood structures and, therefore, must be recalculated. The typical method used to construct a neighborhood in an atomistic system is the radius graph, where neighbors comprise all atoms within a sphere of radius R from a given atom. However, this approach presents a challenge due to the nature of the greater-than-or-equal-to function, which acts as a step function in determining neighbor status. Since the step function is not differentiable, the computational graph would be disconnected leading to poor learning. While employing some kind of smoother function (e.g., sigmoid function) might seem like a natural solution, it remains unclear how to implement message passing on a continuous scale in this context without taking all atoms as neighbors. This in turn would be problematic both in terms of scaling the system and applying the methodology to periodic systems because the computational cost would be prohibitively expensive. To address this issue, we treat the atomic neighborhood as a parameter derived from the corresponding AIMD calculations (see Supplementary Note 1). Importantly, this problem only exists during the training phase as we do not require differentiability during inference.

\subsection{DT accuracy and efficiency}
We validate the proposed DT method on a challenging dataset of AIMD simulations. The dataset contains 100 DFT-based microcanonical trajectories integrated with a timestep $\Delta t=0.5$~fs, resulting in a total of 228,925 atomic configurations. In these simulations, a H$_2$ molecule with an initial translational energy of 0.125~eV interacts with a substrate equilibrated at 300~K. The substrate consists of a Pd$_6$ cluster anchored to a vacancy of a graphene layer. Different processes such as H$_2$ scattering after one or multiple bouncing events, as well as H$_2$ adsorption and H$_2$ dissociation on the Pd$_6$ cluster that often involve Pd$_6$ isomerization are observed~\cite{alducin2019dynamics}. Using these data, we have trained several DT-EGNN models that basically differ in the size of the employed training sets (number of training points and $S_\text{max}$ values). All these models served us to test and confirm the accuracy and efficiency of the DT method as follows. 
First we show that increasing the subsequence length during the training process increases the accuracy of the DT-EGNN models. Next, we demonstrate that increasing the subsequence length also reduces the errors in atomic configurations not seen during the training process, underscoring the data efficiency of the method. 

\begin{figure}[h]
    \centering
    \includegraphics[width=0.7\textwidth]{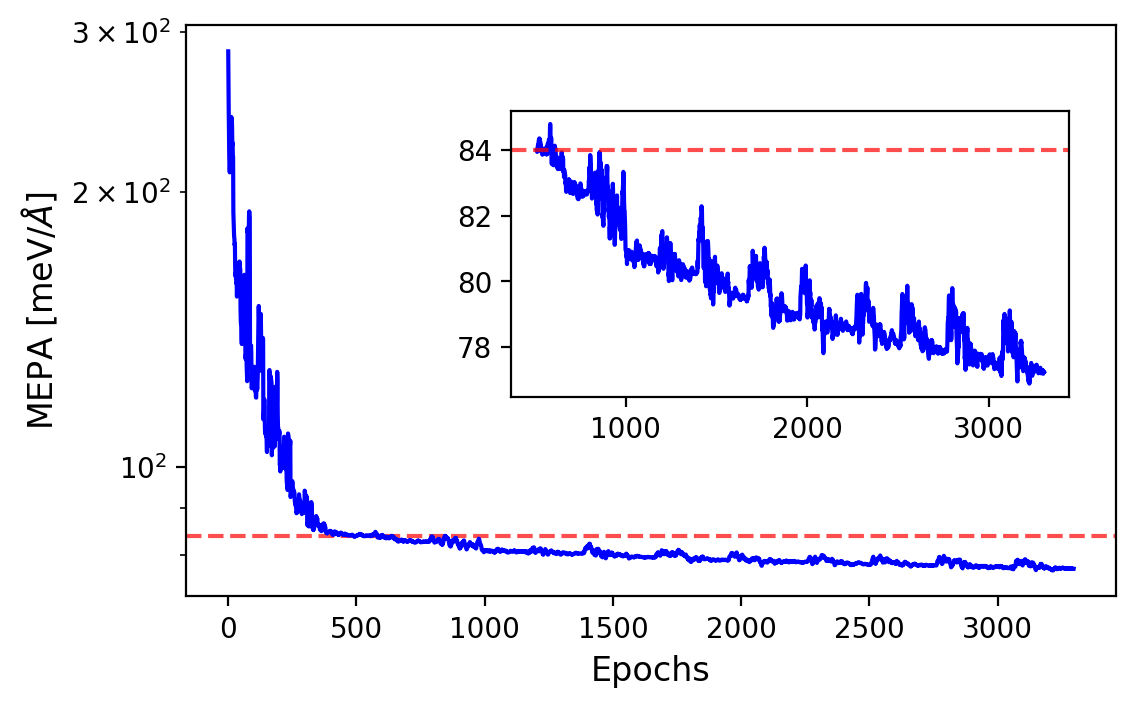}
    \caption{\textbf{Evolution of validation error. } Mean error of atomic forces per atom type and per simulation step (MEPA) in logarithmic scale obtained in the validation set as a function of the training epoch (blue line). 
    Red dashed line indicates the MEPA value in the validation set at first convergence ($S=1$, epoch number 549). The model was trained on 226,825 points with maximum subsequence length $S_\text{max} = 11$. The subsequence length in the validation set is $S_\text{val} = 60$. Inset: Detailed view of the validation error (linear scale) behaviour after the first convergence ($S=1$).}
    \label{fig:accuracy}
\end{figure}

Common practices often involve using validation points that closely resemble training points. In this work however, we adopt a different strategy. Our training set consists of 226,825 points and each training point has a maximum subsequence length $S_{\text{max}}=11$, whereas the validation set contains 2048 points, each having a subsequence length $S_\text{val} = 60$. In other words, each time we evaluate our model against the validation set we perform 2048 MD simulations for 60 integration steps. It is important to point out that while the subsequence length of a training point changes during the training process every time a convergence is reached (see Figure~\ref{fig:flowcharts}\textbf{b}), it stays fixed for a validation point. This provides us with a constant benchmark throughout the training process, while also giving us the opportunity to guide the training process towards a model that best aligns with the demands of a long-lasting molecular dynamics simulation. The validation error curve during the training process of one of the calculated DT-EGNN models for H$_2$/Pd$_6$@G$_{\text{vac}}$ is shown in Figure~\ref{fig:accuracy}. The employed error metric, defined in eq~\eqref{MEPA} in the Methods section and abbreviated as MEPA, represents the mean error of atomic forces per atom type and per simulation step. 
The validation curve exhibits the typical steep decrease that is obtained at the beginning of the training process. The scheduler decreases the learning rate when the validation error shows no improvement over a patience period. When the patience period is exceeded at the minimum learning rate, we consider the model converged for the current subsequence length. First such convergence is marked by the red dashed line in Figure~\ref{fig:accuracy}. Once this criterium is met, the subsequence length is increased by one, and the learning rate reset. This reset value turned out to be a very important hyperparameter because a too small value might cause the neural network to stay trapped in the local minimum it found itself after the last convergence, whereas a too large value might destabilize the learning process. Details of all hyperparameters used in the training processes are provided in the Methods section. The novel ingredient in the DT method is the 
information on the system dynamics that is incorporated through the subsequence length hyperparameter. The benefit of adding such information is confirmed in Figure~\ref{fig:accuracy}, in which we observe that the validation error decreases with the increase in the subsequence length.  However, this decrease is not strictly monotonic due to the fluctuations which largely align with the increase in learning rate following convergence.
Since the validation error represents the average error per simulation step, these validation error reductions have an amplified significance. 

\begin{figure}
    \centering
    \includegraphics[width=0.65\textwidth]{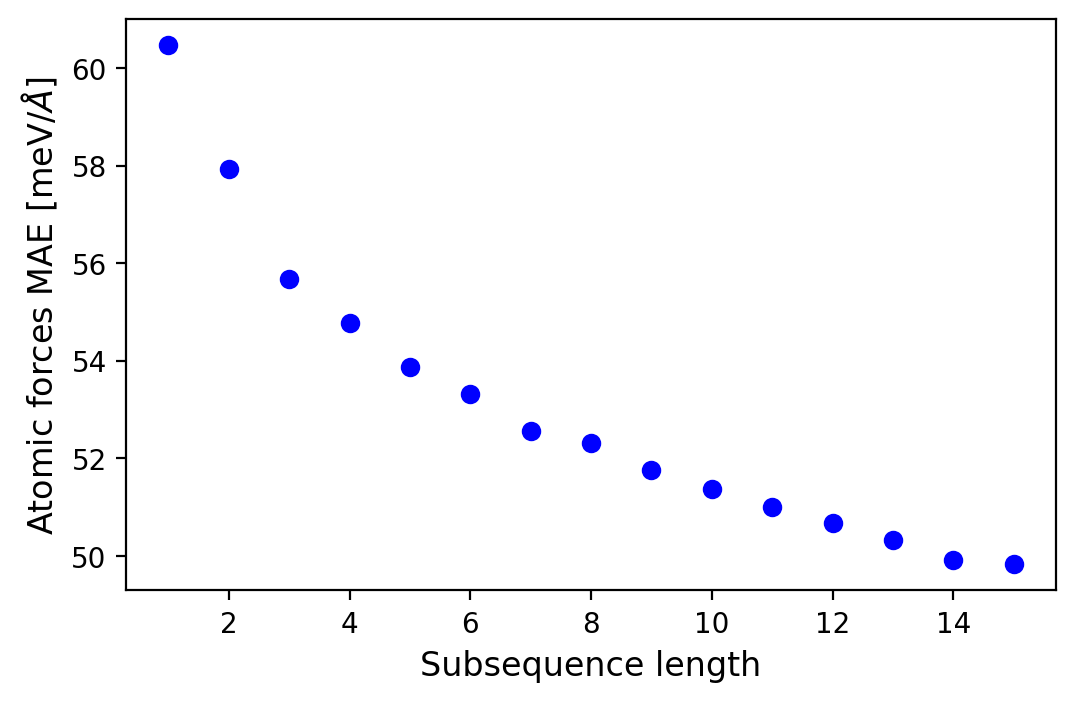}
    \caption{\textbf{DT-EGNN performance with training subsequence length.} Mean absolute error of atomic forces on previously unseen atomic configurations as a function of training subsequence length $S$. At each $S$, the unseen configurations (256 points, each with subsequence length of 120 steps) are evaluated using the model with the lowest validation error at this $S$. The training set consists of 108,019 points, each with a maximum subsequence length $S_\text{max} = 15$ and the validation set contains 512 points, each with a subsequence length $S_\text{val} = 120$.} 
    \label{fig:efficiency}
\end{figure}
A key aspect of any machine learning method is its efficiency in utilizing training data. This is especially true for atomistic systems where ab-initio data are scarce and expensive to come by. We have investigated how well our approach generalizes to atomic structures unseen in the training set. Leaving the model architecture unchanged, we have selected for training a subset of 50 simulations from the original 100 AIMD simulations. This amounts to 108,019 training points with the maximum subsequence length set to $S_{\text{max}}$ = 15. The validation set consists of 512 points with a subsequence length equal to $S_\text{val} = 120$. The 512 points used for validation are chosen from the same subset of 50 simulations used for training so that the other half of the AIMD simulations remain completely invisible to the model. In order to examine how the training subsequence length affects the model performance, we save the model with the lowest validation error at each subsequence length during the training process. Thus, as $S_{\text{max}}$ = 15,  we end up with fifteen models with each model representing the best performing model for its respective training subsequence length. Using a set of 50 previously unseen AIMD simulations, we randomly sampled 256 atomic structures. Each structure served as a starting point for a 120-timesteps simulation, which is then compared to the reference AIMD data. The resulting mean absolute errors (MAE) in atomic forces for each of the fifteen models are shown in Figure~\ref{fig:efficiency}. We observe a consistent decrease in the prediction error with increasing subsequence length, despite the number of unique atomic structures in the training set staying constant throughout the training process.

\begin{figure}
    \centering
    \includegraphics[width=0.65\textwidth]{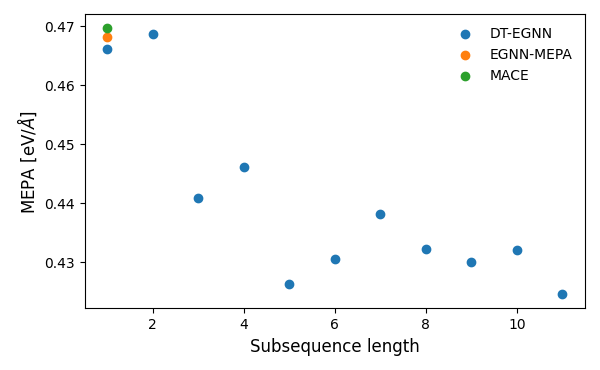}
    \caption{\textbf{DT-EGNN compared to conventionally trained NNPs.} Mean error of atomic forces per atom type and per simulation step (MEPA) as a function of training subsequence length $S$, obtained with the DT-EGNN showcased in Figure~\ref{fig:accuracy}. 
    At each $S$, 100 test MD simulations of 300 integration steps are
evaluated using the model with the lowest validation error at this $S$.
    For comparison, the MEPA obtained by equivalent MD simulations performed with MACE and EGNN-MEPA, which were trained on the complete data set of 100 AIMD simulations, are shown at subsequence length equal to one.}
    \label{fig:final_comparison}
\end{figure}

\subsection{Comparison to conventional training methods}

So far we have examined the properties and performance of the DT approach in isolation. However, to establish its practical value, we compare it to conventional NNP training methods. We evaluate the DT-EGNN model trained using the complete AIMD data set with $S_\text{max}=11$ and $S_\text{val}=60$ against two NNPs, namely MACE \cite{batatia2022mace} and EGNN-MEPA, trained on a complete AIMD data set but in a conventional fashion. These two NNPs provide valuable benchmarks for comparison. EGNN-MEPA has identical architecture and number of parameters to DT-EGNN, differing only in the training method, while MACE represents a powerful, widely-used equivariant graph neural network with a computationally heavier architecture. To test performance beyond the training simulation length of the DT-EGNN models, we selected one random atomic structure from each of the 100 trajectories, resulting in 100 test configurations. From each test configuration and for each NNP, we performed 300-timesteps MD simulations using the AIMD timestep of 0.5~fs. The atomic forces obtained at each integration step are compared to their AIMD counterparts and the corresponding MEPA values are shown in Figure~\ref{fig:final_comparison}. As expected, DT-EGNN at subsequence length equal to one and EGNN-MEPA perform similarly since the only difference at that subsequence length between these two NNPs is the validation scheme of DT-EGNN. However, increasing the subsequence length leads to a clear reduction of MEPA values, demonstrating the benefits of DT even for molecular simulations significantly longer than those encountered during training. Moreover, DT-EGNN outperforms the state-of-the-art NNP MACE, indicating an important trade-off in NNP design. Complex architectures like MACE promise better extrapolation capabilities through built-in symmetry constraints. However, this advantage comes with a high computational cost during inference. Since inference time typically constitutes the larger share of computational expenses compared to training time, DT offers a compelling alternative. It acts as a regularization technique imposed by temporal correlations present in molecular dynamics data and increases an accuracy of a NNP without increasing the complexity of a NNP itself. It might, however, seem surprising that EGNN-MEPA and DT-EGNN (at subsequence length equal to one) have better performance than MACE. This can, at least partly, be explained in terms of the choice of loss function. By default, MACE uses the weighted mean squared error (MSE) of energies and forces as a loss function (eq~\eqref{MSE_loss}). However, such a loss function underperforms when systems contain disproportionate numbers of atoms across different atom types (see Supplementary Note 2). This problem is further exacerbated because hydrogen atoms, despite being the least abundant species with only two atoms present, cover the largest phase space volume in our system. Results of all models are summarized in Table~\ref{table:performance}.

\begin{table}[htbp]
\begin{tabular}{p{0.40\linewidth}c} 
\hline
\textbf{NNP}  & \textbf{MEPA [eV/\AA]}  \\
\hline
EGNN-MEPA  & 0.468 \\
DT-EGNN ($S=1$)   & 0.466 \\
MACE  & 0.470\\
DT-EGNN ($S=11$)  & \textbf{0.424} \\
\hline
\end{tabular}
\caption{Mean error of atomic forces per atom type and per simulation step (MEPA) between NNPs and DFT calculated on 300-timesteps simulations.}
\label{table:performance}
\end{table}

\section{Conclusion}
In conclusion, we have introduced dynamic training, a unique methodology to increase accuracy of NNPs for long-lasting molecular dynamics. Our approach extends beyond the conventional single-point training paradigm by incorporating sequences of consecutive atomic configurations that provide information on the system dynamics. We have validated the method on a challenging dataset of 100 ab-initio molecular dynamics simulations of $\text{H}_{2}$ impinging on
a substrate consisting of a Pd$_6$ cluster anchored to a graphene vacancy. With the help of this dataset we have explored the relation between the accuracy of a model and the training subsequence length. We found that increasing the subsequence length during the training process does indeed improve the accuracy of the DT-EGNN. Furthermore, we have demonstrated the ability of the DT-EGNN method to generalize on atomic structures not present in the training set. Finally, comparing the DT-EGNN to conventionally trained NNPs on simulation lengths much larger than those present in the training set, we show that DT-EGNN outperforms the EGNNs with identical architecture but trained in conventional fashion. Moreover, comparison with state-of-the-art NNP MACE reveals a significant advantage of DT. It enhances NNP accuracy without increasing architectural complexity, offering a computationally efficient alternative to complex architectures that impose higher inference costs. This demonstrates that leveraging temporal correlations through DT provides a practical pathway to improving molecular dynamics simulations while maintaining favorable computational performance.

A key feature of the DT method is that it is agnostic with respect to the architecture of the model. This means that it can be applied to any other already existing NNP. Looking at the relation between the EGNN models trained by the conventional methods and DT-EGNN, it is reasonable to expect that applying the DT method to MACE would also cause an improvement in the performance relative to MACE trained with the conventional method. Another promising direction for future research lies in exploring benefits of incorporating multiple different systems in the training set with varying integration times.

We expect that the proposed method will enable researchers to conduct more accurate and efficient molecular dynamics simulations, particularly in systems in which generating training data is computationally expensive. The demonstrated improvement in data efficiency could significantly impact applications in computational chemistry and materials science, where access to high-quality training data often constitutes a bottleneck in model development.

\section{Methods}\label{sec3}

\subsection{Equivariant Graph Neural Networks}
Graph Neural Networks (GNNs) are a class of deep learning models designed to process data represented as graphs. A graph is a pair $\mathcal{G} = (\mathcal{V}, \mathcal{E})$ where $\mathcal{V}$ represents a set of nodes, and $\mathcal{E}$ set of edges. This structure maps naturally to atomistic systems, where atoms and their interactions can be directly represented within the graph framework. In this work nodes are represented by a one-hot vector denoting the chemical element of an atom, while interatomic distances represent edges between the nodes. Each prediction of a model starts with the embedding layer acting on node and edge features. Let $\mathbf{x}_{0} \in \mathbf{R}^a$ be a one-hot vector representing the atom species, where $a$ is the number of unique atom species in the system, and $\text{d}_{ij}$ a scalar representing the interatomic distance between atoms $i$ and $j$. Then embedding mappings are 
\begin{align}
\phi_n &: \mathbf{R}^a \rightarrow \mathbf{R}^{d_n} \\
\phi_e &: \mathbf{R}^1 \rightarrow \mathbf{R}^{2 d_e},
\end{align}
where $d_n$ and $d_e$ are embedding dimensions for node and edge,  respectively. $\phi_n$ is represented by a multilayer perceptron, and $\phi_e$ is a random Fourier feature mapping \cite{rahimi2007random} of the form,
\begin{equation}
    \phi_e\left( \text{d}_{ij} \right) = \left[\text{sin}\left( \text{b}_1 \text{d}_{ij}\right) , \text{cos}\left( \text{b}_1 \text{d}_{ij}\right),\dots,\text{sin}\left( \text{b}_{d_e} \text{d}_{ij}\right) , \text{cos}\left( \text{b}_{d_e} \text{d}_{ij}\right) \right],
\end{equation}
where each parameter $\text{b}_i$ is sampled from the normal distribution $\mathcal{N}(0,\,\sigma^{2})$. Work by Tancik et al.~\cite{tancik2020fourier} shows that such a mapping can overcome the spectral bias~\cite{basri2020frequency, rahaman2019spectral} inherent to multilayer perceptrons. Embedding layers are followed by message passing layers defined by the following transformations
\begin{align}
    \mathbf{m}_{ij}^l &= \Phi_l\left(\mathbf{h}^l_i, \mathbf{h}^l_j, \phi_e(d_{ij})\right) \\
    \mathbf{m}_i^l &= \sum_{j \in {N}(i)} \mathbf{m}_{ij}^l \\
    \mathbf{h}_i^{l+1} &= \Phi'_l\left(\mathbf{h}_i^l, \mathbf{m}_i^l\right) \, ,
\end{align}
where $\mathbf{h}_{i}^l,\mathbf{m}_{i}^l \in \mathbf{R}^{d_n}$ are the i-th atom node and edge vectors at layer $l$. The neighborhood of an atom $i$ denoted as ${N}(i)$ is calculated by a radius graph. At each layer $l$, update functions $\Phi_l$, and $\Phi'_l$ are represented by multilayer perceptrons. After K message passing layers and global pooling, the final vector $\mathbf{h}^K$ is passed through a final multilayer perceptron $\psi$ to obtain a prediction of the potential energy of the system
\begin{equation}
    E_{\text{pred}} = \psi\left( \mathbf{h}^K  \right) \,.
\end{equation}
Finally, adiabatic atomic forces are obtained by taking the gradient of the predicted system energy

\begin{equation}
    \mathbf{F}_{i} = - \mathbf{\nabla}_i\,E_{\text{pred}} \, .
\end{equation}

\subsection{Training details}
As shown schematically in Figure~\ref{fig:flowcharts}, the training of the potential energy surface starts by the traditional strategy of considering the different atomic structures in the dataset independently, without taking advantage of the dynamical information. In other words, each training data point has information on a single atomic structure and, in the context of our DT method, we denote them as subsequences of length one ($S=1$). At this step, only the atomic positions, energies, and forces for each structure present in the dataset are needed and used for training. More precisely, for this initial subsequence length $S=1$, the loss function in DT-EGNN is defined as
%
\begin{equation}
L\left(S=1 \right) = L_{\text{energy}} + L_{\text{force}} \, ,
\label{loss_S1}
\end{equation}
where
\begin{equation}
    L_{\text{energy}} = \frac{1}{B} \sum_b^B \, \frac{1}{N_b}\left| E_{\text{pred},b} - E_{\text{DFT},b} \right|
    \label{loss_energy}
\end{equation}
and
\begin{equation}
L_{\text{force}} = \frac{1}{B} \sum_b^B \sum_{a \in A_b} \left(  \frac{1}{3 N_{a,b} } \, \sum_{i=1}^{N_{a,b}} \sum_{\alpha = 1}^3 \left| F_{\text{pred},\alpha,b}^{a,i} - F_{\text{DFT},\alpha, b}^{a,i} \right| \right)\,. \label{loss_force}
\end{equation}
Here $B$ is the batch size; $E_{\text{pred},b}$ and $E_{\text{DFT},b}$ are the potential energies of the atomic configuration $b$ calculated by NNP and DFT, respectively; $N_b$ is the total number of atoms in the atomic configuration $b$;  
$A_b$ is the set of the different atomic species present in $b$, with $N_{a,b}$ being the number of atoms of the atomic species $a$ from $A_b$ in atomic configuration $b$.

After the convergence criteria is met for $S=1$, the subsequent length is increased by one. In this case ($S=2$), each data training point has information on two atomic structures, the initial one of the subsequence and the ensuing structure in the corresponding AIMD simulation. More precisely, the quantities extracted from the AIMD simulations 
and used during training are: the atomic positions and velocities of the first structure of the subsequence, the integration time step, and the forces for the two structures of each subsequence. Starting with the positions and velocities of the first structure of the subsequence, the model predicts the corresponding forces that, together with the integration time step, are used to generate the positions and velocities of the next step and predict the corresponding forces. The same procedure is used for larger subsequence lengths. In other words, only the time step, positions and velocities of the first structure of the subsequence are taken from the AIMD dataset. The positions and velocities of the rest of the structures in the subsequence are calculated using the velocity Verlet algorithm, and the forces of all the structures are predicted by the model.
The training is performed by evaluating the accuracy of the model in predicting the forces of all the structures of the subsequence. More precisely, for subsequence lengths larger than one, the loss function is the following:

%
\begin{equation}
L\left(\text{S}> 1 \right) = \sum_{k=1}^{\text{S}} \lambda_{k} \, L_{\text{force}}^{k} \, ,
\end{equation}
where the force term for each subsequence length $L_{\text{force}}^{k}$ is of the same form as in eq~\eqref{loss_force} and $\lambda_{k}$ is equal to 50 for $k$ equal to one, and unity for any $k$ greater than one. Such scaling scheme proved to be crucial for the successful implementation of the method. This importance stems from the fact that during training process, initial structures are the only structures for which atomic positions exactly match those from the corresponding AIMD simulations. Note that, as described in Figure~\ref{fig:flowcharts}, the increase in the subsequence length is performed progressively. It is increased by one each time the force error convergence is met, until the predefined $S_{\text{max}}$ value is reached.

Mean error of atomic forces per atom type and per simulation step (MEPA) is defined as
\begin{equation} \label{MEPA}
\text{MEPA}=\frac{1}{T}\sum_{i=1}^T L_{\text{force}}^{i} \, ,
\end{equation}
where $T$ is the total number of simulation steps and $L_{\text{force}}^{i}$ has the same form as in eq~\eqref{loss_force}.

The weighted mean square error loss function, used to train the MACE NNP, has the following form
\begin{equation} \label{MSE_loss}
    L_{\text{MSE}} = \lambda_E \, \frac{1}{B} \sum_{b}^{B} \frac{1}{N_b} (E_{\text{pred},b} - E_{\text{DFT},b})^2 +  \lambda_F \, \frac{1}{B} \sum_{b}^{B} \frac{1}{3N_b} \sum_{i=1}^{N_b} \sum_{\alpha=1}^3 (F_{\text{pred},\alpha,b} - F_{\text{DFT},\alpha,b})^2 \, ,
\end{equation}
where $\lambda_E$ and $\lambda_F$ are weights for the energy and force term respectively.

All EGNN NNPs used in this work have identical architecture with the following model hyperparameters. Embedded node feature vectors have dimension of 128, whereas embedded edge feature vectors have dimension of 512. The standard deviation of the normal distribution from which random Fourier features were computed is equal to four. There are three message passing layers and the cutoff radius determining the neighborhood structure is equal to 5~\AA. In total, each EGNN in this work has 513,281 learnable parameters. 
 
The MACE NNP used in this work has the following model hyperparameters. Number of invariant and equivariant messages was set to 128. Cutoff radius was set to 5~\AA. The rest of the MACE hyperparameters were set to default values as per mace-torch version 0.3.6. In total, the MACE model in this work has 751,888 learnable parameters.

All NNPs were trained on NVIDIA A100 GPUs in Python under version 3.12, Pytorch under version 2.4.0, and Pytorch Geometric under version 2.5.3. For the models trained with conventional methods (MACE, EGNN-MEPA) we used 95\%/5\% splits resulting in 217,478 atomic configuration in the training set and 11,447 in the validation set. The learning rate, initially set to $ 10^{-3}$, was controlled by the Pytorch's ReduceLRonPlateau scheduler. All NNPs were trained until the minimum learning rate of $2 \times 10^{-6}$ was reached except for the MACE NNP that was trained for 450 epochs. This constitutes nearly four times as many gradient updates than what is recommended in the official MACE documentation as a heuristic. Average epoch training times for the different NNPs employed in this study are summarized in Table \ref{tab:nnp_epoch}. The reset value of the learning rate for DT-EGNN was set to $10^{-4}$. Training batch size was equal to 128 for the NNPs trained in a single-point fashion and 32 for DT-EGNN. Validation batch size was equal to 256 and test batch size was equal to one for all NNPs used in this work. All NNPs used Adam \cite{kingma2014adam} as an optimizer.  

\begin{table}[htbp]
  \begin{tabular}{p{0.40\linewidth}c}
    \hline
    \textbf{NNP} & \textbf{Epoch time [s]} \\
    \hline
     MACE & 3573 \\
     DT-EGNN (S=11) & 1566 \\
     DT-EGNN (S=1) & 196 \\
     EGNN-MEPA &     625    \\
    \hline
  \end{tabular}
  \caption{Average epoch duration (in seconds) for the NNPs used in this work.  DT-EGNN models were trained using 4 GPUs, whereas MACE and EGNN-MEPA were trained using a single GPU.}
  \label{tab:nnp_epoch}
\end{table}

\section{Data and Software Availability}\label{sec4}
AIMD simulations and related informations used in the work have been deposited in the Figshare database under accession code \href{https://doi.org/10.6084/m9.figshare.28498778.v1}{https://doi.org/10.6084/m9.figshare.28498778.v1}. An open-source software implementation of DT-EGNN approach is available at \\ \href{https://github.com/IZugec/DTEGNN}{https://github.com/IZugec/DTEGNN}

\section{Author contributions}\label{sec6}
\textbf{Ivan Žugec} conceptualization, data curation, formal analysis, investigation, methodology, software, validation, writing-original draft, writing-review \& editing; \textbf{Tin Hadži Veljković} methodology, validation, writing-review \& editing; \textbf{Maite Alducin} methodology, validation, supervision, writing-review \& editing, funding acquisition; \textbf{J. Iñaki Juaristi} methodology, validation, supervision, writing-review \& editing, funding acquisition.

\section{Notes}\label{sec7}
The authors declare no competing financial interest. 

\begin{acknowledgement}

Financial support was provided by the
Spanish MCIN$/$AEI$/$10.13039$/$501100011033$/$, FEDER Una manera de hacer Europa (Grant No.~PID2022-140163NB-I00), Gobierno Vasco-UPV/EHU (Project No.~IT1569-22), and the Basque Government Education Departments’ IKUR program, also co-funded by the European NextGenerationEU action through the Spanish Plan de Recuperación, Transformación y Resiliencia (PRTR). Computer resources were provided by the Donostia International Physics Center (DIPC) Supercomputing Center.

\end{acknowledgement}



\appendix

\section*{Supplementary Note 1}
In order to benefit from performing dynamics during the training process, every operation from the first model prediction up to weight adjustments has to be differentiable. This presents a problem when building the atomic neighborhoods within the $S$-loop shown in Fig.~1\textbf{b} in the main text that stems from the binary nature of operations typically used to construct it. One way to deal with this problem is to extract neighborhood information of each atomic structure from the underlying AIMD simulations and record it during the data preprocessing step. Given sufficiently accurate predictions of atomic forces, the neighborhood structure of updated atomic positions will match the one from corresponding AIMD simulations. This is why it is very important to converge the model on subsequence length equal to one before introducing larger lengths in the training process.

Storing neighborhood structures at each step did not exceed our computational limits. For this reason, this is the strategy that we followed in this work. The neighborhood structure was updated at every step according to the corresponding neighborhood structure in the AIMD simulations.

\begin{figure}[h]
\centering
\includegraphics[width=0.8\textwidth]{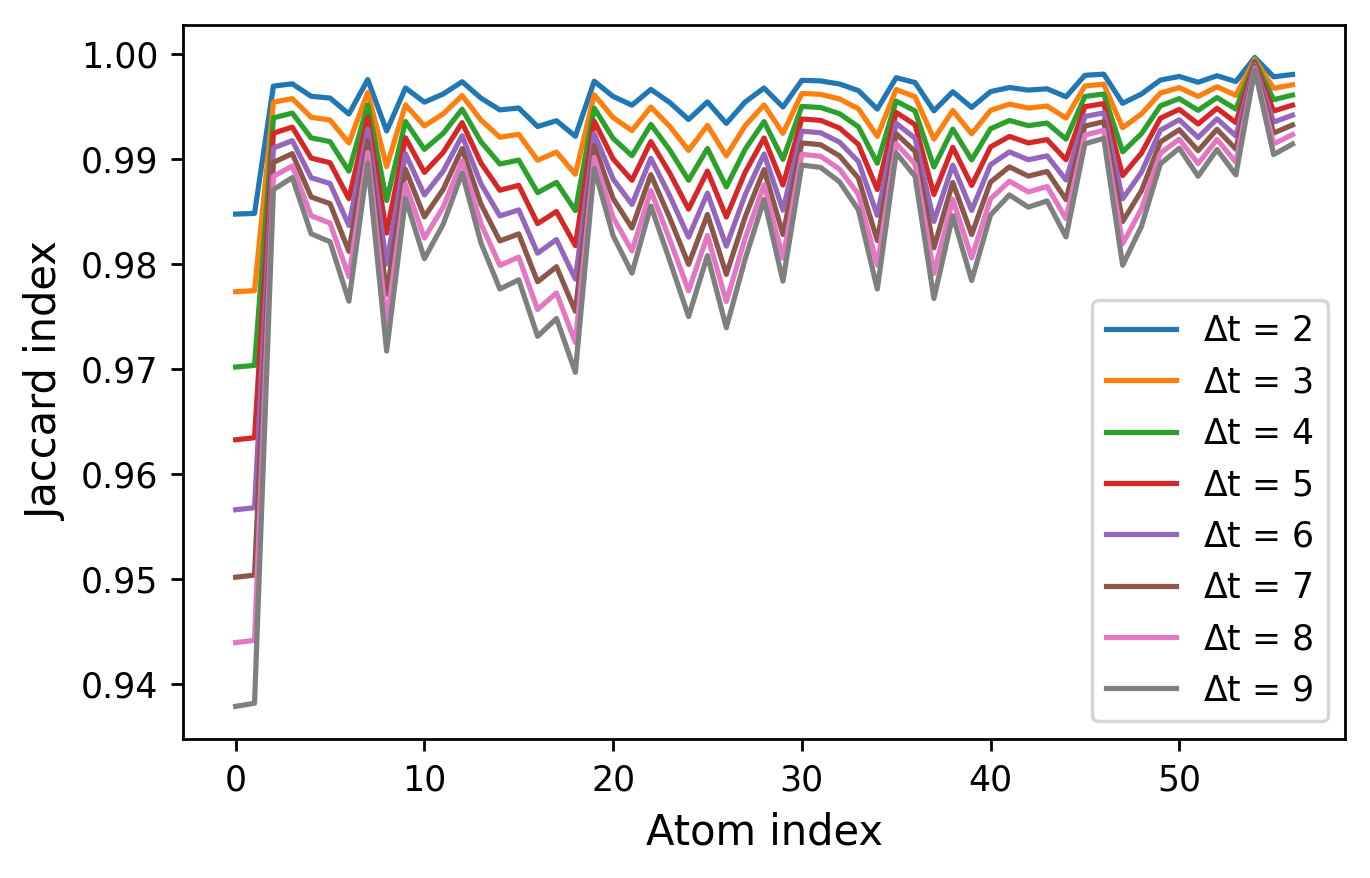}
\caption{Jaccard index of neighborhood structure as a function of atom index in the $\text{H}_2\text{Pd}_6@\text{G}_\text{vac}$ for various offsets $\Delta t$. Indices 0 and 1 correspond to hydrogen atoms forming the $\text{H}_2$ molecule, indices from 2 to 50 correspond to carbon atoms, and indices 51 to 56 correspond to paladium atoms. }
\label{fig:jaccard}
\end{figure}

However, applying the DT method to large datasets and systems with large amount of atoms could increase the computational cost. In other words, for too large systems and too large datasets, storing the AIMD neighboorhood structures in order to update the neighboorhood structure at every step of the $S$-loop could be prohibitive. This is why we propose the following 
idea to alleviate the computational burden. If we assume that the changes in the neighborhood structure occur gradually along the dynamics, we can reuse the same neighborhood structure for multiple simulation steps. To examine the validity of this approximation for the system used in this work, we measure the rate of change of the neighborhood structure for atomic configurations offset by a different amount of simulation steps $\Delta t$. As a measure of a difference between sets of neighbors we use the Jaccard similarity index
\begin{equation}
    J\left({N}(i,t), \, {N}(i,t +\Delta t )\right) = \frac{|{N}(i,t) \cap {N}(i,t +\Delta t )|}{|{N}(i,t) \cup {N}(i,t +\Delta t )|},
\end{equation}
where ${N}(i,t)$ is the set of neighbors of atom $i$ at simulation time $t$. The closer $\Delta t$ is to zero, the closer the Jaccard index is to unity. The averaged Jaccard index for each atom over all 100 AIMD simulations of the training set is shown in Fig.~\ref{fig:jaccard}. Indices 0 and 1 correspond to hydrogen atoms forming the $\text{H}_2$ molecule, indices from 2 to 50 correspond to carbon atoms, and indices 51 to 56 correspond to palladium atoms. As we would expect, hydrogen atoms have the lowest similarity index as they move the most. However, note that after nine simulation steps the averaged similarity is still above 93\% for hydrogen atoms and above 97\% for the rest of atoms in the system. Moreover, training models with neighborhood updates taking place every five steps yielded comparable performance. This framework can therefore serve as a systematic method to determine appropriate neighborhood-structure-update frequencies across different systems in the dataset.

The results of Fig.~\ref{fig:jaccard} show that in the case of our system, instead of updating the atomic neighborhoods at every step, we could have opted for updating them after several steps. They also suggest that this could be the ideal strategy in case of too large systems for which updating at every step would be prohibitive. Note that, the main reason for this is that in well converged AIMD simulations the neigborhood structure varies smoothly. For example, the hydrogen atoms in the aforementioned AIMD simulations explore very different situations and neighborhoods (forming an isolated H$_2$ molecule far from the Pd cluster, forming a H$_2$ molecule interacting strongly with the Pd atoms, dissociated H atoms interacting with the Pd atoms). Still, the variation of the neighborhood structure for these H atoms along the dynamics is smooth enough to allow for updating neighborhood structure after several steps. This constitutes a strong indication of the validity of the proposed strategy for more general and larger systems of atoms.

\section*{Supplementary Note 2}
\begin{figure}[h]
\centering
\includegraphics[width=0.7\textwidth]{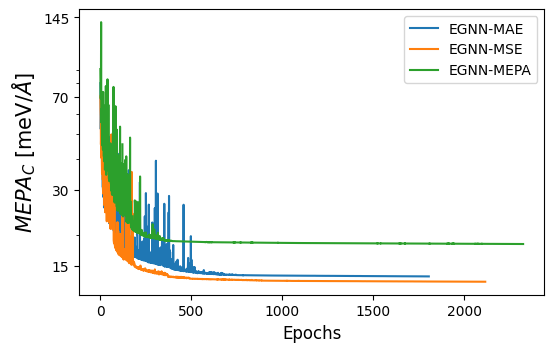}
\caption{Contribution to MEPA coming from carbon atoms as a function of epochs, calculated from atomic structures present in the validation set. All models, namely, EGNN-MAE (blue), EGNN-MSE (orange), and EGNN-MEPA (green), were trained on 217,478 training structures and validated on 11,447  structures.}
\label{fig:2}
\end{figure}

\begin{figure}[h]
\centering
\includegraphics[width=0.7\textwidth]{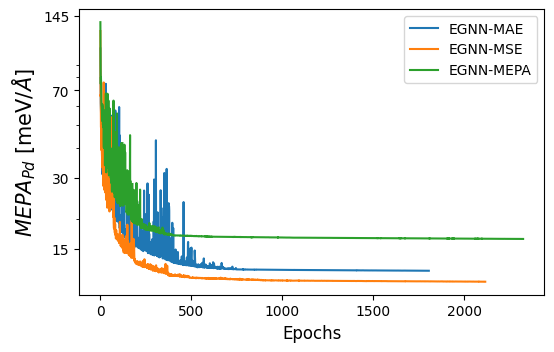}
\caption{Contribution to MEPA coming from palladium atoms as a function of epochs, calculated from atomic structures present in the validation set. All models, namely, EGNN-MAE (blue), EGNN-MSE (orange), and EGNN-MEPA (green), were trained on 217,478 training structures and validated on 11,447  structures.}
\label{fig:3}
\end{figure}

\begin{figure}[h]
\centering
\includegraphics[width=0.7\textwidth]{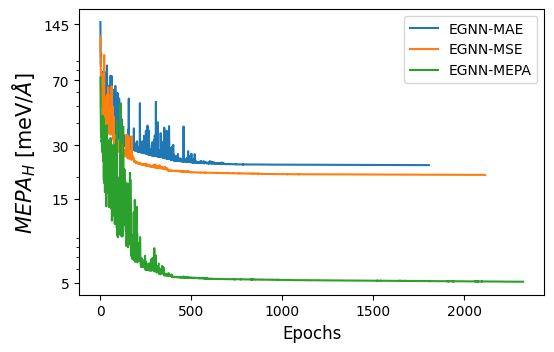}
\caption{Contribution to MEPA coming from hydrogen atoms as a function of epochs, calculated from atomic structures present in the validation set. All models, namely, EGNN-MAE (blue), EGNN-MSE (orange), and EGNN-MEPA (green), were trained on 217,478 training structures and validated on 11,447 structures.}
\label{fig:4}
\end{figure}

Here we show how the choice of the loss function directly impacts the NNP performance during extended molecular dynamics simulations. The comparison involves three models: EGNN-MEPA, EGNN-MAE, and EGNN-MSE named after the loss functions used during training process. Weighted MSE loss function, used to train MACE as well as EGNN-MSE, and MEPA are already defined in the Methods section of the main text. Mean absolute error (MAE) loss function is defined as 
\begin{equation} \label{MAE_loss}
    L_{\text{MAE}} =\frac{1}{B} \sum_{b}^{B} \frac{1}{N_b} \left|E_{\text{pred},b} - E_{\text{DFT},b}\right| +  \frac{1}{B} \sum_{b}^{B} \frac{1}{3N_b} \sum_{i=1}^{N_b} \sum_{\alpha=1}^3 \left|F_{\text{pred},\alpha,b} - F_{\text{DFT},\alpha,b}\right| .
\end{equation}
To ensure a controlled comparison, all NNPs have identical architectures and equal number of trainable parameters. Moreover, all of them were trained on the same training and validation sets. Figs.~\ref{fig:2}--\ref{fig:4} show the components of MEPA during the training process on structures present in the validation set for each atomic species (carbon, palladium, and hydrogen) present in the system. Even though the validation error given to the scheduler during the training process was consistent with the loss function for each respective model, we calculate the MEPA for each model in order to compare them. Models trained with MAE and MSE loss functions show better performance on carbon and palladium atoms, but perform significantly worse for hydrogen atoms. This is because MAE and MSE focus on minimizing the errors of the whole structure, while MEPA focuses on minimizing the error of each atom species. 

\clearpage
\section*{Supplementary Note 3}
Here we demonstrate the ability of DT-EGNN to capture a complex dynamical feature of the $\mathrm{H}_2\mathrm{Pd}_6@\mathrm{G}{\mathrm{vac}}$ system that typically emerges only after thousands of integration steps. In Ref. \citenum{alducin2019dynamics} it was shown that hydrogen‐dissociation events are often accompanied by a structural rearrangement of the palladium cluster. More precisely, the dissociation of H$_2$ in the Pd cluster can induce its transition from the original octahedral structure to an incomplete pentagonal bypiramid structure. It was shown that the structural transition basically consists in the drastic elongation of the Pd-Pd distance between two particular Pd neighbors in the equatorial plane of the octahedron. It is worth mentioning that this structural change is a relatively long time process that takes place 1-3~ps after H$_2$ dissociation.

\begin{figure}[htbp]
  \centering
  \begin{subfigure}[b]{0.7\textwidth}
    \centering
    \includegraphics[width=\linewidth]{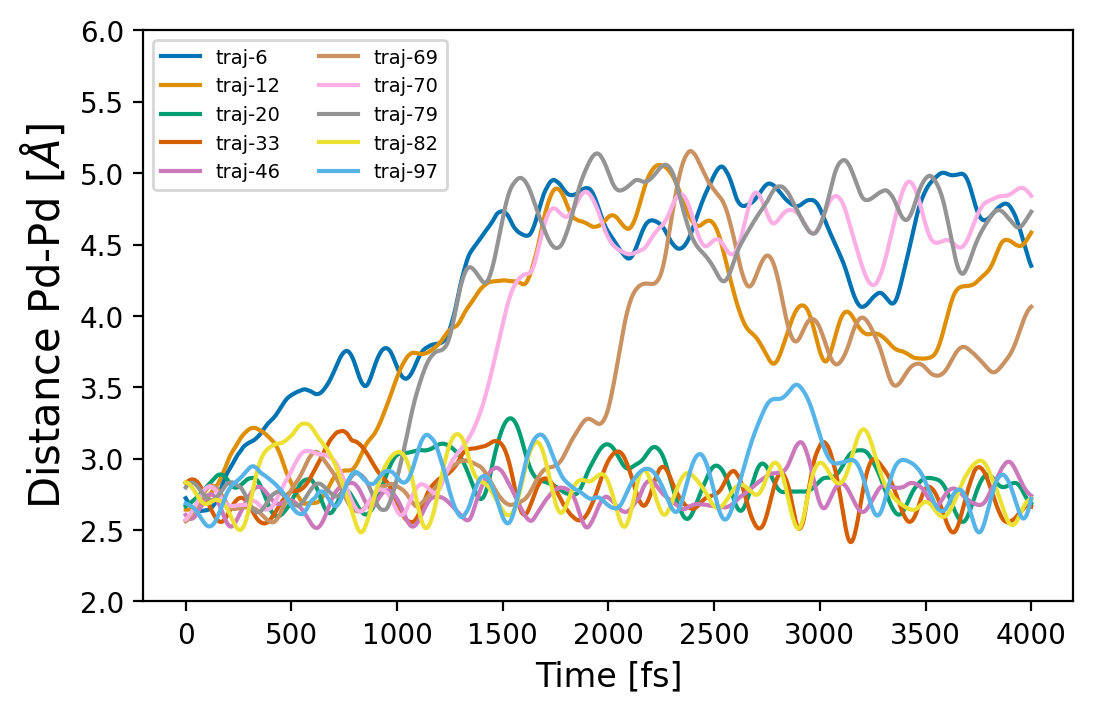}
  \end{subfigure}
  \vspace{1em}
  \begin{subfigure}[b]{0.7\textwidth}
    \centering
    \includegraphics[width=\linewidth]{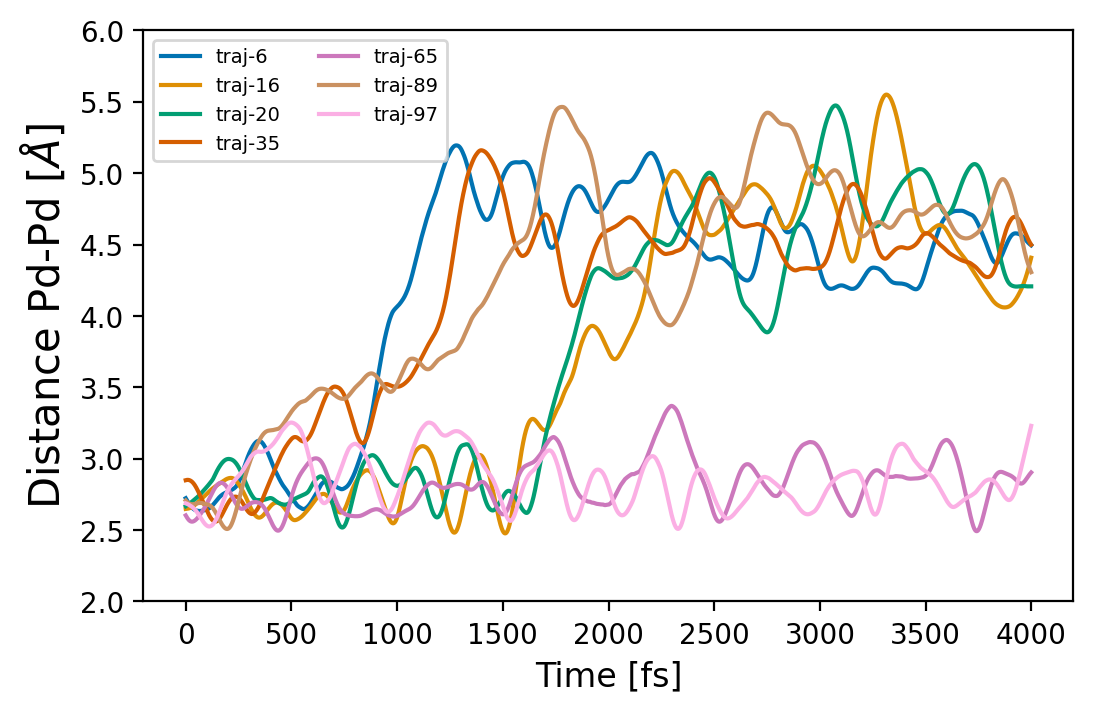}
  \end{subfigure}
  \caption{Distance between the two Pd atoms exhibiting pronounced elongation along the trajectories corresponding to dissociative adsorption of H$_2$ for DT-EGNN NNP (top) and AIMD (bottom).}
  \label{fig:isomerization}
\end{figure}

In order to test whether we can observe such events, we used the same initial conditions as the corresponding AIMD calculations and, using DT-EGNN as PES, propagated those initial conditions for 4~ps with the integration time step equal to 0.1~fs. Note that this requires performing 40~000 integration steps, to be compared with the 11 steps utilized in the longest training subsequences. The obtained results suggest that not only do we observe events featuring dissociation of the hydrogen molecule, but also reproduce the structural transition of the palladium cluster. Fig~\ref{fig:isomerization} shows the time evolution of the distance between the two relevant palladium atoms for both the DT-EGNN NNP and the AIMD reference. Although a sample of 100 trajectories is insufficient for statistically meaningfull discussion of dissociation probabilities (10 events for DT-EGNN vs. 7 for AIMD), it is noteworthy that three of these events (trajectories 6, 20 and 97) share identical initial conditions. All in all, these results show the good performance of the method on long time dynamics and its ability to describe complex dynamical phenomena.

\clearpage

\section*{Supplementary Note 4}

Here we summarize the information on all DFT and computational settings that were employed in Ref.~\citenum{alducin2019dynamics} to run the AIMD simulations, from which we obtained the dataset used in the present manuscript.

The calculations were based on spin-polarized DFT for a H$_2$ molecule impinging on a Pd$_6$ cluster anchored to a graphene vacancy (the system denoted as Pd$_6$@G$_{\text{vac}}$ in the main text). The plane waves based VASP code~\cite{kresse96,kresse94} was used to perform the calculations and exchange correlation effects were described using the generalized gradient approximation PW91 functional~\cite{perdew92}. The interaction of the explicitly considered valence electrons with the atomic cores was treated in the projector augmented wave (PAW) approximation~\cite{bloch94}, for which the PAW potentials supplied with the VASP package were used~\cite{kresse99}. The energy cutoff for the plane-wave basis set was 400~eV. The integration in the Brillouin zone was performed using a $\Gamma$-centered 2$\times$2$\times$1 Monkhorst-Pack grid of special \textbf{k}-points~\cite{monkhorst76}. The first-order Methfessel-Paxton broadening scheme, with a 0.1~eV width, was used to consider fractional electronic-state occupancies.~\cite{methfessel89}

The supercell consisted of a hexagonal 5$\times$5 graphene layer in the lateral directions. The height of the supercell in the normal direction to the graphene layer was 14~{\AA}. The substrate was relaxed until the forces on each Pd and C were below 0.02~eV/{\AA}. 

The ab initio molecular dynamic simulations used to generate the neural network potential energy surface in the present work were obtained for a H$_2$ molecule impinging on Pd$_6$@G$_{\text{vac}}$ under normal incidence conditions, with an initial substrate temperature of 300~K. The zero pressure limit was studied by considering one H$_2$ per simulated trajectory.  The initial translational and vibrational energies of the H$_2$ molecule were E$_i=0.125$~eV and E$_{vib}$($\nu =$ 0, j$=$0) = 0.27~eV, respectively. The center of mass of the molecules was initially located at a height 9~{\AA}above the graphene layer. The initial lateral position of the center of mass and the orientation of the molecule were randomly sampled. Before running the AIMD simulations, the Pd$_6$@G$_{\text{vac}}$ substrate was equilibrated at 300~K for 7~ps using the Nos\'{e} thermostat~\cite{Nose84}. Sunsequently, the energy-conserving AIMD simulations of the H$_2$ molecules impinging on the anchored cluster were performed, taking the initial positions and velocities of the C and Pd atoms randomly from the set of configurations generated
during the thermalization at $T=300$~K. The number of calculated trajectories was 100, the integration time step 0.5~fs, and the total integration time was 4~ps.

\section*{Supplementary Note 5}
Here, we show that our DT-EGNN NNP can perform molecular dynamics simulations lasting several orders of magnitude longer than those encountered during training, while maintaining the stability of the system. Additionally, we provide direct evidence for improved stability as the subsequence length, used during training process, increases.

\begin{figure}[H]
\centering
\includegraphics[width=0.7\textwidth]{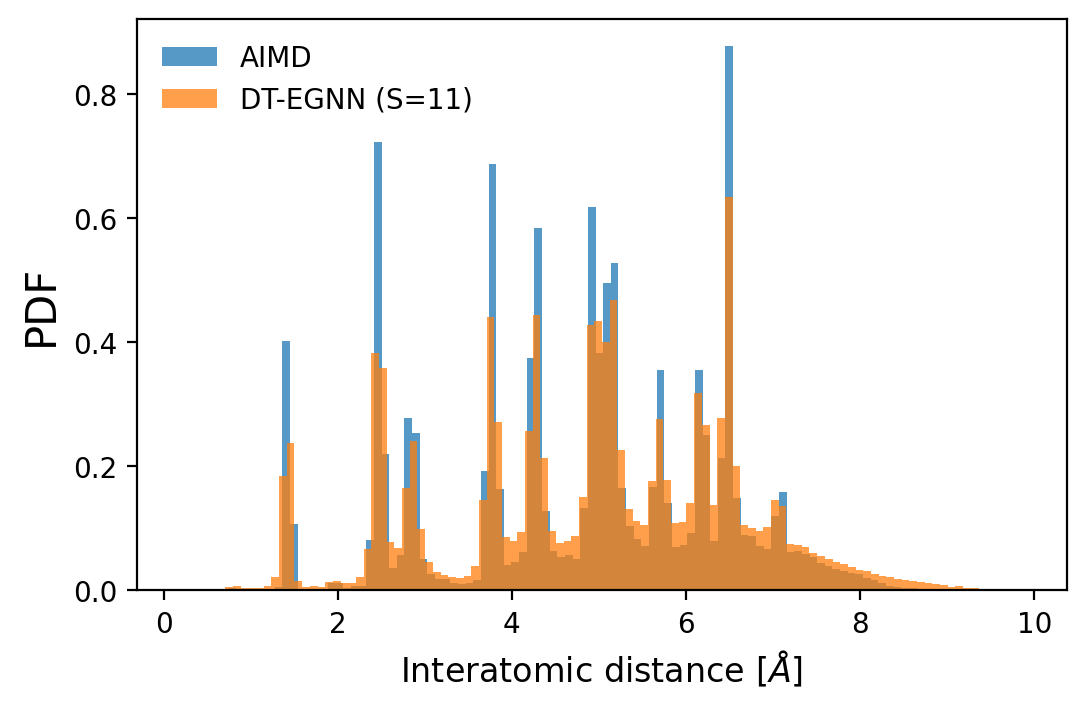}
\caption{Comparison of normalized histograms of interatomic distances extracted from 100 independent MD trajectories generated by DT-EGNN ($S=11$, orange, snapshots at 100 ps) and 100 AIMD trajectories (blue, durations 800 fs–4 ps).}
\label{fig:100ps}
\end{figure}

Using DT-EGNN ($S=11$) as a NNP, 100 MD simulations with the integration time step of 0.5~fs and total time of 100~ps were performed. Taking the last snapshot of each of the 100 trajectories, we calculate distances between all atoms in the system and present them as a histogram. Furthermore, for comparison, we do the exact same thing for 100 simulations performed with AIMD. A comparison of normalized densities of interatomic distances between these two methods is shown on Fig.~\ref{fig:100ps}.

It is important to stress that the histogram for DT-EGNN is created for the atomic configurations at the end of 100~ps, while AIMD simulations last between 800~fs and 4~ps. Still, almost all of the peaks present in the AIMD structures are also present after 100~ps, providing compelling evidence that our DT-EGNN approach reliably maintains structural stability well beyond the subsequence lengths used during training.

\begin{figure}[H]
\centering
\includegraphics[width=0.7\textwidth]{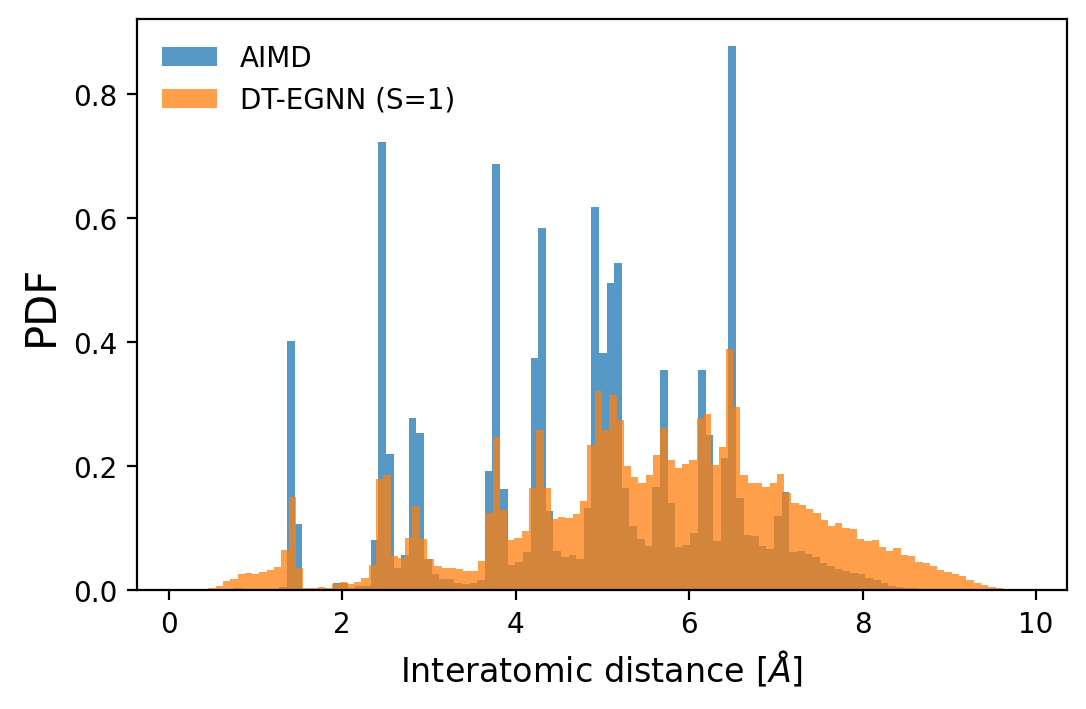}
\caption{Comparison of normalized histograms of interatomic distances extracted from 100 independent MD trajectories generated by DT-EGNN ($S=1$, orange, snapshots at 100 ps) and 100 AIMD trajectories (blue, durations 800 fs–4 ps).}
\label{fig:100ps_S1}
\end{figure}

Finally, we also performed 100 MD simulations using DT-EGNN at subsequence length equal to one ($S=1$) and compared the resulting distribution of interatomic distances at 100~ps to those obtained from AIMD simulations (see Fig.~\ref{fig:100ps_S1}). Relative to DT-EGNN at $S=11$, results for DT-EGNN at $S=1$ suggest that the discrete peaks of interatomic distances are less obvious for this distribution. Furthermore, significant portion of interatomic distances lay outside of the range of the AIMD distribution (very small as well as very big interatomic distances) suggesting much stronger deviations from the initial structure.

\bibliography{achemso-demo}
\end{document}